\documentstyle[12pt,a4,epsf]{article}
\begin{document}
\bibliographystyle{unsrt}
\title{Statistical Mechanics in a Nutshell\thanks{Part I
of a course on ``Transport Theory'' taught at
Dresden University of Technology, Spring 1997.
For more information see the course web site\newline
{\tt http://www.mpipks-dresden.mpg.de/$\sim$jochen/transport/intro.html}}
}
\author{Jochen Rau\\
Max-Planck-Institut f\"ur Physik komplexer Systeme\\
N\"othnitzer Stra{\ss}e 38, 01187 Dresden, Germany}
\maketitle
\tableofcontents
\clearpage

\section{Some Probability Theory}\label{probability}

\subsection{Constrained distributions}

A random experiment has $n$ possible results at each trial;
so in $N$ trials there are $n^N$ conceivable outcomes.
(We use the word ``result'' for a single trial, while ``outcome''
refers to the experiment as a whole; thus one outcome consists of
an enumeration of $N$ results, including their order.
For instance, ten tosses of a die ($n=6,N=10$) might have
the outcome ``1326642335.'')
Each outcome yields a set of sample numbers $\{N_i\}$ and
relative frequencies $\{f_i=N_i/N, i=1\ldots n\}$.
In many situations the outcome of a random experiment is not known completely:
One does not know the order in which the individual results occurred, and
often one does not even know all $n$ relative frequencies $\{f_i\}$ but
only a smaller number $m$ ($m<n$) of linearly independent constraints
\begin{equation}
\sum_{i=1}^n G^i_a f_i = g_a
\quad,\quad
a=1\ldots m
\, .
\label{constraint}
\end{equation}

As a simple example consider a loaded die.
Observations on this badly balanced die have shown that $6$ occurs
twice as often as $1$; nothing peculiar was observed for the other
faces.
Given this information only and nothing else, i.e., not making
use of any additional information that we might get from
inspection of the die or from past experience with dice in general,
all we know is a single constraint of the form (\ref{constraint})
with
\begin{equation}
G_1^i =\left\{
\begin{array}{rl}
2: & i=1 \\
0: & i=2\ldots 5 \\
-1: & i=6
\end{array}
\right.
\label{die}
\end{equation}
and $g_1=0$.

The available data --in the form of linear constraints--
are generally not sufficient to reconstruct
unambiguously the relative frequencies $\{f_i\}$.
These frequencies may be regarded as Cartesian coordinates of 
a point in an $n$-dimensional vector space.
The $m$ linear constraints, together with $f_i\in[0,1]$ and the 
normalization
condition $\sum f_i=1$, then just restrict the allowed points to
some portion of an $(n-m-1)$-dimensional hyperplane.

\subsection{Concentration theorem}

Given an a priori probability distribution $\{p_i\}$
for the results $i=1\ldots n$,
the probability that $N$ trials will yield the
--generally different-- relative frequencies $\{f_i\}$ is
\begin{equation}
\mbox{prob}(\{f_i\}|\{p_i\},N)
=
{N!\over N_1!\ldots N_n!}\,
p_1^{N_1}\ldots p_n^{N_n}
\quad.
\end{equation}
Here 
the second factor is the probability for one specific outcome
with sample numbers $\{N_i\}$,
and the first factor counts
the number of all outcomes that give rise to the
same set of sample numbers. 
With the definition
\begin{equation}
{I}_p(f):=-\sum_i f_i \ln {f_i\over p_i}
\end{equation}
and the shorthand notations $f=\{f_i\}$, $p=\{p_i\}$
we can also write
\begin{equation}
\mbox{prob}(f|p,N)
=
\mbox{prob}(f|f,N)\exp[N{I}_p(f)]
\quad.
\end{equation}
In particular,
for two different data sets $\{f_i\}$ and $\{f'_i\}$ the ratio
of their respective probabilities is given by
\begin{equation}
{\mbox{prob}(f|p,N)\over \mbox{prob}(f'|p,N)}=
{\mbox{prob}(f|f,N)\over \mbox{prob}(f'|f',N)}\,
\exp[N({I}_p(f)-{I}_p(f'))]
\end{equation}
where, by virtue of Stirling's formula
\begin{equation}
x!\approx \sqrt{2\pi x}\, x^x e^{-x}
\quad,
\end{equation}
it is asymptotically
\begin{equation}
{\mbox{prob}(f|f,N)\over \mbox{prob}(f'|f',N)}
\approx
\sqrt{\prod_i{f'_i\over f_i}}
\quad.
\end{equation}
As the latter ratio is independent of $N$,
for large $N$ and nearby distributions $f'\approx f$
the variation of $\mbox{prob}(f|p,N)/\mbox{prob}(f'|p,N)$ 
is completely dominated by
the exponential:
\begin{equation}
{\mbox{prob}(f|p,N)\over \mbox{prob}(f'|p,N)}\approx
\exp[N({I}_p(f)-{I}_p(f'))]
\quad.
\label{ways}
\end{equation}
Hence the probability with which any given frequency
distribution $f$ is realized is essentially determined by the
quantity ${I}_p(f)$:
The larger this quantity, the more likely the 
frequency distribution is realized.

Consider now all frequency distributions allowed by $m$
linearly independent constraints. 
As we discussed earlier, the allowed distributions can be
visualized as points in
some portion of an $(n-m-1)$-dimensional hyperplane.
In this hyperplane portion there is a unique point at which
the quantity ${I}_p(f)$ attains a maximum $I_p^{\rm max}$;
we call this point 
the ``maximal point'' $f^{\rm max}$.
(That the maximal point is indeed unique 
can be seen as follows: Suppose there were not one but two
maximal points corresponding to frequency distributions
$f^{(1)}$ and $f^{(2)}$. Then the mixture
$\bar{f}=(f^{(1)}+f^{(2)})/2$ would have 
${I}_p(\bar{f})>I_p^{\rm max}$, which would be a contradiction.)
It is possible to define new coordinates 
$\{x_1\ldots x_{n-m-1}\}$ in the hyperplane such that
\begin{itemize}
\item
they are linear functions of the $\{f_i\}$;
\item
the origin ($\vec{x}=0$) is at the maximal point; and
\item
in the vicinity of the maximal point
\begin{equation}
I_p(\vec{x})=I_p^{\rm max}-a r^2 + O(r^3)
\quad,\quad
a>0
\quad,
\end{equation}
where
\begin{equation}
r:=\sqrt{\sum_{j=1}^{n-m-1} x_j^2}
\quad.
\end{equation}
\end{itemize}
Frequency distributions that satisfy the given constraints 
(\ref{constraint}) and whose $I_p(\vec{x})$ differs from
$I_p^{\rm max}$ by {more} than $\Delta\! I$
thus lie outside
a hypersphere around the maximal point,
the sphere's radius $R$ being given by $aR^2=\Delta I$.
The probability that $N$ trials will yield such
a frequency
distribution outside the hypersphere is 
\begin{equation}
\mbox{prob}(I_p<(I_p^{\rm max}-\Delta\! I)
|m\, \mbox{constraints})
=
{\int_R^\infty dr\,r^{n-m-2}\exp(-Nar^2) \over
\int_0^\infty dr'\,{r'}^{n-m-2}\exp(-Na{r'}^2)}
\quad.
\end{equation}
Here the factors $r^{n-m-2}$ in the integrand are due to the
volume element, while the exponentials
$\exp(-Nar^2)=\exp(N(I_p(\vec{x})-I_p^{\rm max}))$ 
stem from the ratio (\ref{ways}).
Substituting 
$t=Nar^2$, defining
\begin{equation}
s:=(n-m-3)/2
\end{equation}
and using
\begin{equation}
\Gamma(s+1)=\int_0^{\infty} dt\,t^s\exp(-t)
\end{equation}
one may also write
\begin{equation}
\mbox{prob}(I_p<(I_p^{\rm max}-\Delta\! I)
|m\, \mbox{constraints})
= {1\over\Gamma(s+1)} {\int_{N\Delta I}^\infty dt\,t^s\exp(-t)}
\quad;
\end{equation}
which for large $N$ ($N\gg s/\Delta I$) can be approximated
by
\begin{equation}
\mbox{prob}(I_p<(I_p^{\rm max}-\Delta\! I)
|m\, \mbox{constraints})
\approx
{1\over\Gamma(s+1)} (N\Delta I)^s \exp(-N\Delta I)
\, .
\label{approx}
\end{equation}
As the number $N$ of trials increases, this probability
rapidly tends to zero
for any finite $\Delta I$. 
As $N\to\infty$, therefore, it becomes virtually certain
that the (aside from $m$ constraints) unknown
frequency distribution has an $I_p$ very close to $I_p^{\rm max}$.
Hence not only does the maximal point
represent the frequency distribution that is the most likely to
be realized (cf. Eq. (\ref{ways}));
but in addition, as $N$ increases, 
all other --theoretically allowed--
frequency distributions become more and more
concentrated near this maximal point.
Any frequency distribution other than the maximal point
becomes highly atypical of those allowed by the constraints.

\subsection{Frequency estimation}\label{estimation}

We have seen that the knowledge of $m$ ($m<n$) 
``averages'' (\ref{constraint})
constrains, but fails to specify uniquely, the 
relative frequencies $\{f_i\}$.
In view of this incomplete information the relative
frequencies must be {\em estimated}.
Our previous considerations suggest that
the most reasonable estimate is the maximal point:
that distribution which, while satisfying
all the constraints, maximizes $I_p(f)$.
This leads to
a variational equation
\begin{equation}
\delta\left[\sum_i f_i \ln {f_i\over p_i} +\eta\sum_{i} f_i +
\sum_{a} \lambda^a\sum_{i} G^i_a f_i \right]=0
\end{equation}
where the constraints, as well as the normalization condition
$\sum_{i} f_i = 1$, have been implemented
by means of Lagrange multipliers. 
Its solution is of the form
\begin{equation}
f_i^{\rm max}={1\over Z}\exp\left(\ln p_i
-\langle\ln p\rangle_p -\sum_a\lambda^a G_a^i\right)
\label{gen_canonical}
\end{equation}
with 
\begin{equation}
Z=\sum_i \exp\left(\ln p_i-\langle\ln p\rangle_p 
-\sum_a\lambda^a G_a^i\right)
\quad.
\end{equation}
The term
\begin{equation}
\langle\ln p\rangle_p := \sum_j p_j \ln p_j
\end{equation}
has been introduced by convention;
it cancels from the ratio in (\ref{gen_canonical})
and so does not affect the frequency estimate.
The expression in the exponent simplifies if and only if
the a priori distribution $\{p_i\}$ is uniform:
In this case,
\begin{equation}
\ln p_i - \langle\ln p\rangle_p = 0
\quad.
\end{equation}
The $m$ Lagrange parameters $\{\lambda^a\}$ must be adjusted such as
to yield the correct prescribed averages $\{g_a\}$.
They can be determined from 
\begin{equation}
{\partial\over\partial\lambda^a}\ln Z
= -g_a
\quad,
\label{lagrange}
\end{equation}
a set of $m$ simultaneous equations for $m$ unknowns.
Finally, inserting (\ref{gen_canonical})
into the definition of $I_p(f)$ gives
\begin{equation}
I_p^{\rm max}=\langle\ln p\rangle_p 
+ \ln Z+\sum_a \lambda^a g_a 
\quad.
\end{equation}
 
There remains the task of specifying 
the --possibly nonuniform-- a priori probability 
distribution $\{p_i\}$.
The $\{p_i\}$ are those probabilities one would assign 
{\em before} having asserted the existence of
the constraints (\ref{constraint}); i.e.,
being still in a state of ignorance.
This ``ignorance distribution'' can usually be determined on the basis
of symmetry considerations:
If the problem at hand is {\em a priori} invariant under
some characteristic group 
then the $\{p_i\}$, too, must exhibit this same
group invariance.\footnote{The rationale underlying this
consistency requirement has historically been called 
the ``Principle of Insufficient Reason'' 
(J. Bernoulli, {\em Ars Conjectandi}, 1713).}
For example, if a priori 
we do not know anything about the properties of a given die
then our prior ignorance
extends to all faces equally. The problem is therefore 
invariant under a relabelling of the faces, which
trivially implies
$\{p_i=1/6\}$. In more complicated random experiments, especially those
involving continuous and hence coordinate-dependent
distributions,
the task of specifying the a priori distribution
may be less straightforward.\footnote{see for example E. T. Jaynes,
{\em Prior probabilities}, IEEE Trans. Systems Sci. Cyb. {\bf 4},
227 (1968)
}

For illustration let us return to the example of the loaded die,
characterized solely by the single constraint (\ref{die}).
What estimates should we make of the relative frequencies $\{f_i\}$ with
which the different faces appeared?
Taking the a priori probability distribution 
--assigned to the various faces
{\em before} one has asserted the die's imperfection--
to be uniform, $\{p_i=1/6\}$,
the best estimate (\ref{gen_canonical}) for the frequency distribution 
reads 
\begin{equation}
f_i^{\rm max} =\left\{
\begin{array}{rl}
Z^{-1} \exp(-2\lambda^1): & i=1 \\
Z^{-1}: & i=2\ldots 5 \\
Z^{-1} \exp(\lambda^1): & i=6
\end{array}
\right.
\end{equation}
with only a single
Lagrange parameter $\lambda^1$ and 
\begin{equation}
Z=
\exp(-2\lambda^1)+ 4 + \exp(\lambda^1)
\quad.
\end{equation}
The Lagrange parameter is readily determined from
\begin{equation}
{\partial\over\partial\lambda^1}\ln Z
= -g_1 = 0
\quad,
\end{equation}
with solution 
\begin{equation}
\lambda^1=(\ln 2)/3
\quad.
\end{equation}
This in turn gives the numerical estimates
\begin{equation}
f_i^{\rm max} =\left\{
\begin{array}{rl}
0.107: & i=1 \\
0.170: & i=2\ldots 5 \\
0.214: & i=6
\end{array}
\right.
\end{equation}
with an associated
\begin{equation}
I_p^{\rm max}=\ln(1/6) + \ln Z=-0.019
\quad.
\end{equation}

The above algorithm for estimating frequencies can be iterated.
Suppose that beyond the $m$ constraints (\ref{constraint}) we
learn of $l$ additional, linearly independent constraints
\begin{equation}
\sum_{i=1}^n G^i_a f_i = g_a
\quad,\quad
a=(m+1)\ldots (m+l)
\, .
\end{equation}
In order to make an improved estimate that takes 
these additional data into account
we can either,
(i) starting from the same a priori distribution $p$ as before,
apply the algorithm 
to the total set of $(m+l)$ constraints; or
(ii) iterate: use the previous estimate (\ref{gen_canonical}), which
was based on the first
$m$ constraints only, as a new
a priori distribution $f^{\rm max}\mapsto p'$,
and then repeat the algorithm just for the $l$ additional 
constraints.
Both procedures give the {\em same} improved estimate ${f^{\rm max}}'$.
Associated with this improved estimate is
\begin{equation}
{I_p^{\rm max}}'=I_p^{\rm max} + I_{f^{\rm max}}({f^{\rm max}}')
\quad.
\end{equation}

\subsection{Hypothesis testing}

Now we consider random experiments for which complete frequency
data {\em are} available.
Suppose that, based on some insight we have
into the systematic influences affecting
the experiment,
we conjecture that the observed relative frequencies can be
fully characterized by a set of constraints of the 
--by now familiar-- form (\ref{constraint}), 
and that hence the observed relative frequencies can be {\em fitted}
with a maximal distribution (\ref{gen_canonical}).
This maximal distribution contains $m$ fit parameters $\{\lambda^a\}$
(the Lagrange parameters) whose specific values depend on
the averages $\{g_a\}$, which in turn are
extracted from the data.
It represents our {\em theoretical model} or {\em hypothesis}.

In general, the experimental frequencies $f$
and the theoretical fit $f^{\rm max}$
do not agree exactly. Must the hypothesis therefore be rejected, or is
the deviation merely a statistical fluctuation?
The answer is furnished by the concentration theorem:
Let $N$ be
the number of trials performed to
establish the experimental distribution, let
\begin{equation}
\Delta I = I_p^{\rm max} - I_p(f)
\end{equation}
and $s=(n-m-3)/2$.
For large $N$ ($N\gg s/\Delta I$) the probability 
that statistical fluctuations alone yield an $I_p$-difference
as large as $\Delta I$ is given by (\ref{approx});
typically the hypothesis is rejected whenever this probability
is below $5\%$,\footnote{The hypothesis test presented here 
is closely related to the
better-known $\chi^2$ test.}
\begin{equation}
\mbox{prob}(I_p<(I_p^{\rm max}-\Delta\! I)
|m\, \mbox{constraints})
<5\%
\quad.
\end{equation}
Rejecting a hypothesis means that
the chosen set of constraints was not complete,
and hence that important systematic effects have been overlooked.
These must be incorporated in the form of additional constraints.
In this fashion one can proceed iteratively from simple to ever
more sophisticated models until the deviation of the fit from
the experimental data ceases to be statistically 
significant.

\subsection{Jaynes' analysis of Wolf's die data}

The above prescription for testing hypotheses and --if 
rejected-- for iteratively improving them by enlarging the
set of constraints has been lucidly illustrated by E. T. Jaynes
in his analysis of Wolf's die data.\footnote{E. T. Jaynes,
{\em Concentration of distributions at entropy maxima}, in:
E. T. Jaynes, Papers on Probability, Statistics and
Statistical Mechanics, ed. by R. D. Rosenkrantz, Kluwer
Academic, Dordrecht (1989).
}
Rudolph Wolf (1816--1893), a Swiss astronomer, had performed a number
of random experiments, presumably to check the validity of
statistical theory.
In one of these experiments a die (actually two dice, but only
one of them is of interest here) was tossed $20,000$ times in a
way that precluded any systematic favoring of any face over any other.
The observed relative frequencies $\{f_i\}$ and their deviations 
$\{\Delta_i=f_i-p_i\}$ from the a priori probabilities
$\{p_i=1/6\}$ are given in Table \ref{die_table}.
Associated with the observed distribution is
\begin{equation}
I_p(f)=-0.006769
\quad.
\end{equation}
\begin{table}
\begin{center}
\begin{tabular}{c|c|c}
$i$ &
$f_i$ &
$\Delta_i$
\\
\hline
1 &
0.16230 &
-0.00437
\\ 
2 &
0.17245 &
+0.00578
\\ 
3 &
0.14485 &
-0.02182
\\ 
4 &
0.14205 &
-0.02464
\\ 
5 &
0.18175 &
+0.01508
\\ 
6 &
0.19960 &
+0.02993
\\ 
\end{tabular}
\caption{Wolf's die data: frequency distribution $f$ and
its deviation $\Delta$ from the uniform distribution.
\label{die_table}
}
\end{center}
\end{table}

Our ``null hypothesis'' H0 is that the die is ideal and 
hence that there are no constraints needed to characterize
any imperfection ($m=0$); the deviation of the experimental
from the uniform distribution, with associated
\begin{equation}
I_p^{\rm max(H0)}=
I_p(p)=0
\quad,
\end{equation}
is merely a statistical fluctuation.
However, the probability that 
statistical fluctuations alone yield an $I_p$-difference
as large as 
\begin{equation}
\Delta I^{\rm H0}= I_p^{\rm max(H0)}- I_p(f) = 0.006769
\end{equation}
is practically zero:
Using Eq. (\ref{approx})
with $N=20,000$ and $s=3/2$ we find
\begin{equation}
\mbox{prob}(I_p<(I_p^{\rm max}-\Delta\! I^{\rm H0})
|0\, \mbox{constraints})
\sim
10^{-56}
\quad.
\end{equation}
Therefore, the null hypothesis is rejected: The die cannot
be perfect.

Our analysis need not stop here. Not knowing
the mechanical details of the die we can still formulate
and test hypotheses as to the
nature of its imperfections.
Jaynes argued that the two most likely imperfections are:
\begin{itemize}
\item
a shift of the center of gravity due to the mass of ivory
excavated from the spots, which being proportional to the
number of spots on any side, should make the ``observable''
\begin{equation}
G_1^i=i-3.5
\end{equation}
have a nonzero average $g_1\ne 0$; and
\item
errors in trying to machine a perfect cube, which will tend
to make one dimension (the last side cut) slightly different
from the other two. It is clear from the data that Wolf's die
gave a lower frequency for the faces (3,4); and therefore that
the (3-4) dimension was greater than the (1-6) or (2-5) ones.
The effect of this is that the ``observable''
\begin{equation}
G_2^i =\left\{
\begin{array}{rl}
1: & i=1,2,5,6 \\
-2: & i=3,4
\end{array}
\right.
\end{equation}
has a nonzero average $g_2\ne 0$.
\end{itemize}
Our hypothesis H2 is that these are the only two imperfections
present. More specifically, 
we conjecture that the observed relative frequencies
are characterized by just two constraints ($m=2$) imposed
by the measured averages 
\begin{equation}
g_1=0.0983
\quad\mbox{and}\quad
g_2=0.1393
\quad;
\end{equation}
and that hence
the observed relative frequencies
can be fitted with a maximal distribution
\begin{equation}
f_i^{\rm max(H2)}={1\over Z}
\exp\left(-\sum_{a=1}^2\lambda^a G_a^i\right)
\quad.
\end{equation}
In order to test our hypothesis we determine
\begin{equation}
Z=\sum_{i=1}^6 \exp\left(-\sum_{a=1}^2\lambda^a G_a^i\right)
\quad,
\end{equation}
fix the Lagrange parameters by requiring
\begin{equation}
{\partial\over\partial\lambda^a}\ln Z
= -g_a
\end{equation}
and then calculate 
\begin{equation}
I_p^{\rm max(H2)}=\ln(1/6) + \ln Z+\sum_{a=1}^2 \lambda^a g_a
\quad.
\end{equation}
With this algorithm Jaynes found 
\begin{equation}
I_p^{\rm max(H2)}=-0.006534
\end{equation}
and thus
\begin{equation}
\Delta I^{\rm H2}= I_p^{\rm max(H2)}- I_p(f) = 0.000235
\quad.
\end{equation}
The probability for such an $I_p$-difference to occur as a 
result of statistical fluctuations is (with now $s=1/2$)
\begin{equation}
\mbox{prob}(I_p<(I_p^{\rm max}-\Delta\! I^{\rm H2})
|2\, \mbox{constraints})
\approx
2.5 \%
\quad,
\end{equation}
much larger than the previous $10^{-56}$ but still below
the usual acceptance bound of 5\%.
The more sophisticated model H2 is therefore a major improvement
over the null hypothesis H0 and captures the principal features
of Wolf's die; yet there are indications
that an additional very tiny imperfection may have been present.

Jaynes' analysis of Wolf's die data furnishes a useful paradigm for the
experimental method in general.
All modern experiments at particle colliders (CERN, Desy, Fermilab\ldots),
for example, yield data in the form of frequency distributions 
over discrete ``bins''
in momentum space, for each of the various
end products of the collision.
The search for interesting signals in the data (new particles,
new interactions, etc.) essentially proceeds in the same manner
in which Jaynes revealed the imperfections of Wolf's die:
by formulating physically motivated hypotheses and testing them
against the data.
Such a test is always statistical in nature.
Conclusions (say, about the presence
of a top quark, or about the presence of a certain imperfection of
Wolf's die) can never be drawn with absolute certainty
but only at some --quantifiable-- confidence level.

\subsection{Conclusion}

In all our considerations a crucial role has been played by
the quantity $I_p$:
The algorithm that yields the best estimate for an unknown
frequency distribution is based on the
maximization of $I_p$; and hypotheses can be tested with the help of
Eq. (\ref{approx}), i.e., by simply comparing
the experimental and theoretical values of $I_p$.
We shall soon encounter the quantity $I_p$ 
again and see how it is related to one
of the most fundamental concepts in statistical mechanics:
the ``entropy.''

\section{Macroscopic Systems in Equilibrium}

\subsection{Macrostate}\label{macrostate}

For complex systems with many degrees of freedom (like
a gas, fluid or plasma) the exact microstate is usually not known.
It is therefore impossible to assign to the system a unique
point in phase space (classical) or a unique wave function
(quantal), respectively.
Instead one must resort to a statistical description:
The system is described by a classical {\em phase space distribution}
$\rho({\bf{\pi}})$ or an
incoherent {\em mixture}
\begin{equation}
\hat{\rho}=\sum_i f_i |i\rangle\langle i|
\end{equation}
of mutually orthogonal
quantum microstates $\{|i\rangle\}$, respectively.
(Where the distinction between classical and quantal does not matter
we shall use the generic symbol $\rho$.)
Probabilities must be real, non-negative, and normalized to one;
which implies the respective properties
\begin{equation}
\rho({\bf{\pi}})^*=\rho({\bf{\pi}})\quad,\quad
\rho({\bf{\pi}})\ge 0\quad,\quad
\int d{\bf{\pi}}\,\rho({\bf{\pi}}) = 1
\end{equation}
or
\begin{equation}
\hat{\rho}^\dagger=\hat{\rho}\quad,\quad 
\hat{\rho}\ge 0\quad,\quad \mbox{tr}\,\hat{\rho}=1
\quad.
\end{equation}
In this statistical description every observable $A$ (real phase space
function or Hermitian operator, respectively) is
assigned an 
{\em expectation value}
\begin{equation}
\langle A\rangle_\rho =
\int d{\bf{\pi}}\,\rho({\bf{\pi}})A({\bf{\pi}})
\end{equation}
or
\begin{equation}
\langle A\rangle_\rho  =
\mbox{tr}(\hat{\rho}\hat{A})
\quad,
\end{equation}
respectively.

Typically, not even the distribution $\rho$ 
is a priori known.
Rather, the state of a complex physical system is characterized by 
very few
macroscopic data. These data may come in
different forms:
\begin{itemize}
\item
as {\em data given with certainty}, such as the type of particles
that make up the system, or the shape and volume of the box in which
they are enclosed.
These exact data we take into account through the definition of the
phase space or Hilbert space in which we are
working;
\item
as {\em prescribed expectation values}
\begin{equation}
\langle G_a\rangle_\rho  = g_a
\quad,\quad a=1\ldots m
\end{equation}
of some set $\{G_a\}$ of selected macroscopic observables. Examples
might be the average total energy, average angular momentum, or 
average magnetization.
Such data, which are of a statistical nature, impose constraints
of the type (\ref{constraint}) on the distribution $\rho$; or
\item
as additional {\em control parameters} on which the selected
observables $\{G_a\}$ may explicitly depend, 
such as an external electric or magnetic field. 
\end{itemize}
According to our general considerations in Section \ref{estimation}
the best estimate for the thus
characterized macrostate is a distribution of the
form (\ref{gen_canonical}).
In the classical case this implies
\begin{equation}
\rho({\bf{\pi}})={1\over Z} 
\exp\left(\ln \sigma({\bf{\pi}}) - \langle\ln\sigma\rangle_\sigma
-\sum_a\lambda^a {G}_a({\bf{\pi}})\right)
\label{cl_macro}
\end{equation}
with
\begin{equation}
Z=\int d{\bf{\pi}}\,
\exp\left(\ln \sigma({\bf{\pi}}) - \langle\ln\sigma\rangle_\sigma
-\sum_a\lambda^a {G}_a({\bf{\pi}})\right)
\quad;
\end{equation}
while for a quantum system
\begin{equation}
\hat{\rho}={1\over Z} \exp\left(\ln\hat{\sigma}
 - \langle\ln\sigma\rangle_\sigma
-\sum_a\lambda^a \hat{G}_a\right)
\label{qu_macro}
\end{equation}
and
\begin{equation}
Z=\mbox{tr}\,\exp\left(\ln\hat{\sigma}
 - \langle\ln\sigma\rangle_\sigma
-\sum_a\lambda^a \hat{G}_a\right)
\quad.
\end{equation}
In both cases $\sigma$ denotes the 
a priori distribution.
The auxiliary quantity $Z$ is referred to as the
{\em partition function}.\footnote{Readers already familiar with
statistical mechanics might be disturbed by the appearance of
$\sigma$ in the definitions of $\rho$ and $Z$. Yet this is essential
for a consistent formulation of the theory:
see, for instance, our remarks at the end of Section
\ref{estimation} on the possibility of
iterating the frequency estimation
algorithm.
In most practical applications $\sigma$ is 
uniform and hence $\ln\sigma-\langle\ln\sigma\rangle_\sigma=0$.
Our definitions of $\rho$ and $Z$ then reduce to the conventional
expressions.}

The phase space integral or trace
in the respective expressions for $Z$ 
depend on the specific choice of the phase space or Hilbert space; 
hence they may depend
on parameters like the volume or particle number.
Furthermore,
there may be an explicit dependence of
the observables $\{G_a\}$ 
or of the a priori distribution $\sigma$
on additional control parameters.
Therefore, the partition function generally depends
not just on the Lagrange multipliers $\{\lambda^a\}$
but also on some other
parameters $\{h^b\}$.
In analogy with the relation (\ref{lagrange})
one then defines new variables
\begin{equation}
\gamma_b:=
{\partial\over\partial h^b}\ln Z
\quad.
\label{gamma}
\end{equation}
(In contrast to (\ref{lagrange}) there is no minus sign.)
The $\{g_a\}$, $\{\lambda^a\}$, $\{h^b\}$ and $\{\gamma_b\}$
are called the {\em thermodynamic variables} of the system;
together they specify the system's macrostate.
The thermodynamic variables
are not all independent: Rather, they are related by (\ref{lagrange})
and (\ref{gamma}), that is, via partial derivatives of $\ln Z$.
One says that $h^b$ and $\gamma_b$, or
$g_a$ and $\lambda^a$, are {\em conjugate} to each other.

Some combinations of thermodynamic variables are of particular importance,
which is why the associated distributions go by
special names.
If the observables that characterize
the macrostate 
--in the form of sharp values given with certainty, or in
the form of expectation values--
are all constants of the motion 
then the system is said to be in {\em equilibrium}.
Associated is an {\em equilibrium
distribution} of the form
(\ref{cl_macro}) or (\ref{qu_macro}),  with
all $\{G_a\}$ being constants of the motion.
Such an equilibrium distribution is itself constant in time,
and so are all
expectation values calculated from it.\footnote{Here we have assumed
that there is no time-dependence of the
 a priori distribution $\sigma$.}
The set of constants of the motion always includes the Hamiltonian 
(Hamilton function or Hamilton operator, respectively)
provided it is not explicitly time-dependent.
If its value for a specific system, the {\em internal energy}, and 
the other macroscopic data are all given with certainty
then the resulting equilibrium distribution
is called {\em microcanonical;}
if just the energy is given on average,
while all other data are given with certainty, {\em canonical;}
and if both energy and total particle number are given on average,
while all other data are given with certainty,
{\em grand canonical}.

Strictly speaking, every description of the macrostate
in terms of thermodynamic variables represents a hypothesis:
namely, the hypothesis that the sets $\{G_a\}$ and  $\{h^b\}$
are actually complete.
This is analogous to Jaynes' model for Wolf's die, which
assumes that just two imperfections (associated with two
observables $G_1,G_2$) suffice to characterize the experimental data.
Such a hypothesis may well be rejected by experiment.
If so, this does not mean that our rationale for constructing
$\rho$ --maximizing $I_\sigma$ under given constraints-- was wrong.
Rather, it means that important macroscopic observables or
control parameters (such as ``hidden'' constants of the motion, or
further imperfections of Wolf's die) have
been overlooked, and that the correct description of the macrostate
requires additional thermodynamic variables.

\subsection{First law of thermodynamics}

Changing the values of 
the thermodynamic variables alters the distribution
$\rho$ and with it the associated 
\begin{equation}
I_\sigma^{\rm max}\equiv I_\sigma(\rho)=
\langle\ln\sigma\rangle_\sigma + \ln Z + \sum_a \lambda^a g_a
\quad.
\label{imax}
\end{equation}
By virtue of Eqs. (\ref{lagrange}) and (\ref{gamma})
its infinitesimal variation is given by
\begin{equation}
{d} I_\sigma^{\rm max}=
d \langle\ln\sigma\rangle_\sigma 
+ \sum_a \lambda^a{d} g_a
+ \sum_b \gamma_b{d} h^b
\quad.
\end{equation}
As the set of constants of the motion always contains the Hamiltonian
its value for the given system, the {internal energy} $U$,
and the associated conjugate parameter, 
which we denote by $\beta$, 
play a particularly important role.
Depending on whether the energy is given with certainty or on average,
the pair $(U,\beta)$ corresponds to a pair $(h,\gamma)$ or $(g,\lambda)$.
For all remaining variables one then defines new 
conjugate parameters
\begin{equation}
l^a:=\lambda^a/\beta\quad,\quad m_a:=\gamma_a/\beta
\end{equation}
such that
in terms of these new parameters the energy differential reads
\begin{equation}
{d} U ={\beta^{-1}{d} (I_\sigma^{\rm max}-\langle\ln\sigma\rangle_\sigma) }
{-\sum_a l^a{d} g_a-
\sum_b m_b{d} h^b}
\quad.
\end{equation}

A change in internal energy that is effected solely by a
variation of the parameters $\{g_a\}$ or $\{h^b\}$ is defined
as {\em work}
\begin{equation}
\delta W:={-\sum_a l^a{d} g_a-
\sum_b m_b{d} h^b}
\quad;
\end{equation}
some commonly used pairs $(g,l)$ and $(h,m)$ of thermodynamic variables
are listed in Table \ref{table_variables}.
If, on the other hand, these parameters are held fixed
(${d} g_a= d h^b=0$) then
the internal energy can still change through the addition
or subtraction of {\em heat}
\begin{equation}
\delta Q:={1\over k\beta}\,k\,
{d}(I_\sigma^{\rm max}-\langle\ln\sigma\rangle_\sigma)
\quad.
\end{equation}
Here we have introduced an arbitrary constant $k$.
Provided we choose this constant to be the
{\em Boltzmann constant}
\begin{equation}
k = 1.381\times 10^{-23} \mbox{J/K}
\quad,
\end{equation}
we can identify the {\em temperature}
\begin{equation}
T := {1\over k\beta}
\end{equation}
and the {\em entropy}
\begin{equation}
S := k\, (I_\sigma^{\rm max}-\langle\ln\sigma\rangle_\sigma)
\end{equation}
to write $\delta Q$ in the more familiar form
\begin{equation}
\delta Q={T{d} S}
\quad.
\end{equation}
The entropy is related to the other thermodynamic variables 
via Eq. (\ref{imax}), i.e.,\footnote{Even though the
entropy, like the partition
function, is
related to measurable quantities it is essentially
an auxiliary concept
and does not
itself constitute a physical observable:
In quantum mechanics, for example, there is nothing like
a Hermitian ``entropy operator.''
}
\begin{equation}
S 
= k\,\ln Z + k \sum_a \lambda^a g_a
\quad.
\label{entropy_relation}
\end{equation}
The relation 
\begin{equation}
dU=\delta Q + \delta W
\quad,
\end{equation}
which reflects nothing but energy conservation, is known as the
{\em first law of thermodynamics}.
\begin{table}
\begin{center}
\begin{tabular}{|c|c|l|}
\hline
$(g,l)$&
$(h,m)$&
names
\\
\hline\hline
&$(V,p)$&volume, pressure
\\ \hline
$(N,-\mu)$&$(N,-\mu)$&particle number, chemical potential
\\ \hline
$(M,-B)$&$(B,M)$&magnetic induction, magnetization
\\ \hline
$(P,-E)$&$(E,P)$&electric field, electric polarization
\\ \hline
$(\vec p,-\vec v)$&&momentum, velocity
\\ \hline
$(\vec L,-\vec\omega)$&&angular momentum, angular velocity
\\ \hline
\end{tabular}
\caption{Some commonly used pairs of thermodynamic variables.
In cases where two pairs are given, e. g., $(M,-B)$ and
$(B,M)$, the proper choice depends on the specific
situation:
For example, the pair $(M,-B)$ is adequate if the magnetization $M$ is a
constant of the motion whose value is given on average;
while the pair $(B,M)$ should be used if there is an externally applied
magnetic field $B$ which plays the role of a control
parameter.
\label{table_variables}
}
\end{center}
\end{table}

\subsection{Example: Ideal quantum gas}

We consider a gas of non-interacting bosons or fermions.
We suppose that the total particle number is {\em not} 
given with certainty (but possibly on average, as in the
grand canonical ensemble) so the system must be
described in Fock space.
We further suppose that the observables $\{\hat{G}_a\}$ 
whose expectation values
are furnished as macroscopic data are all of the single-particle form
\begin{equation}
\hat{G}_a=\sum_i G_a^i \hat{N}_i
\quad,
\end{equation}
where the $\{G_a^i\}$ are arbitrary ($c$-number) coefficients and 
the $\{\hat{N}_i\}$ denote number operators pertaining to some
orthonormal basis
$\{|i\rangle\}$ of single-particle states.
Provided the a priori distribution $\sigma$ is uniform,
the best estimate for the macro\-state
has the form
\begin{equation}
\hat{\rho}={1\over Z}\exp\left(-\sum_i\alpha^i \hat{N}_i\right)
\end{equation}
with 
\begin{equation}
\alpha^i=\sum_a \lambda^a G_a^i
\quad.
\end{equation}
For example, in the grand canonical ensemble
(energy and total particle number
given on average) the parameters $\{\alpha^i\}$ are functions of
the single-particle energies $\{\epsilon^i\}$,
the inverse temperature $\beta$ and the chemical
potential $\mu$:
\begin{equation}
\alpha^i=\beta(\epsilon^i-\mu)
\quad.
\end{equation}
The partition function
\begin{equation}
Z=
\mbox{tr}\,\exp\left(-\sum_{i}\alpha^{i}\hat{N}_i\right)
=
\sum_{{\rm configurations\ } \{N_1, N_2, \ldots \}}
\prod_i \left(e^{-\alpha^{i}}\right)^{N_i}
\end{equation}
factorizes, for we work in Fock space 
where we sum freely over each $N_i$:
\begin{equation}
Z=
\prod_i \sum_{N_i} \left(e^{-\alpha^{i}}\right)^{N_i}
=:
\prod_i Z_i
\quad.
\end{equation}
The sum over $N_i$ extends from $0$ to the maximum value allowed
by particle statistics:
$\infty$ for bosons, $1$ for fermions.
Consequently, each factor $Z_i$ reads
\begin{equation}
Z_i=\left(1\mp e^{-\alpha^{i}}\right)^{\mp 1}
\quad,
\end{equation}
the upper sign pertaining to
bosons and the lower sign to fermions.
This gives
\begin{equation}
\ln Z=\mp\sum_i \ln \left(1\mp e^{-\alpha^{i}}\right)
\end{equation}
and hence the average occupation 
\begin{equation}
n_i \equiv
\langle {N}_i\rangle_\rho= -{\partial\over\partial\alpha^i}\ln Z
= \left(e^{\alpha^i}\mp 1\right)^{-1}
\label{occupation}
\end{equation}
of any single-particle state $i$.
Using the inverse relation
\begin{equation}
\alpha^i = \ln (1\pm n_i)
- \ln n_i
\end{equation}
together with the specific realization of Eq. (\ref{entropy_relation}),
\begin{equation}
S = k\,\ln Z + k \sum_i \alpha^i n_i
\quad,
\end{equation}
we find for the entropy
\begin{equation}
S=-k\sum_i \left[n_i
\ln n_i \mp 
(1\pm n_i)\ln(1\pm n_i)
\right]
\quad.
\end{equation}

\subsection{Thermodynamic potentials}

\begin{figure}
 \centerline{\epsfbox{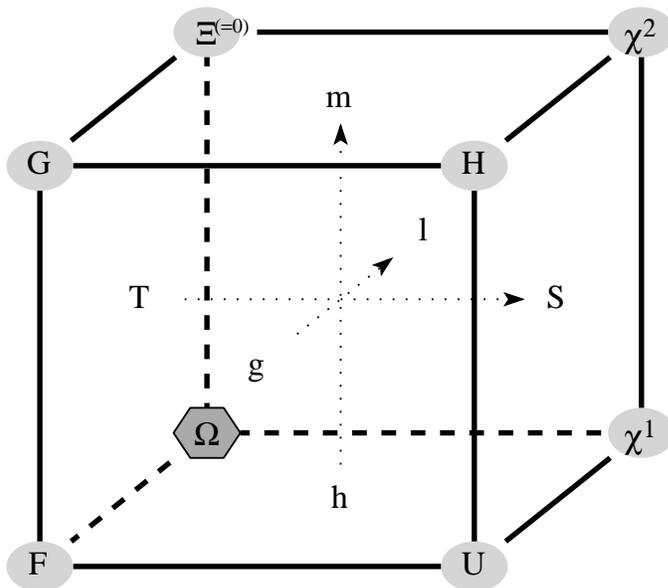}}
\caption{Born diagram.
Corners correspond to thermodynamic
potentials: the grand potential $\Omega$, the free
energy $F$, the internal energy $U$, the enthalpy $H$,
the free enthalpy $G$,
the potential $\Xi$
(which vanishes for a homogeneous system),
as well as two rarely used
potentials $\chi^1$ and $\chi^2$.
Sides of the cube correspond to thermodynamic variables:
$T$, $S$, $g$, $l$, $h$ and $m$.
Opposite sides are conjugate to each other,
and associated with each conjugate pair is a 
dotted ``basis vector.''
Each corner is a function of the adjacent sides;
e.g., the enthalpy $H$ is a function of $\{S,g,m\}$.
Their conjugates $\{T,l,h\}$
can be obtained from $H$
by partial differentiation, the sign depending on
whether the requested conjugate variable is at the head ($-$)
or tail ($+$) of a basis vector;
e.g., $T=+\partial H/\partial S$.
One can go from one corner to the next by moving parallel
or antiparallel to a basis vector, thereby
(i) changing variables such as to get the correct 
dependence of the new potential, and
(ii) adding (if moving parallel) or subtracting (if moving
antiparallel) the product of the conjugate variables that are
associated with the basis vector.
For instance, in order to obtain the free enthalpy $G$ from
the enthalpy $H$ one
(i) uses $T=+\partial H/\partial S$ to solve for
$S(T,g,m)$, since the free enthalpy will be a function of
$\{T,g,m\}$ rather than $\{S,g,m\}$; and then
(ii) subtracts the product $TS$ to get
$G(T,g,m)=H(S(T,g,m),g,m) - TS(T,g,m)$.
This procedure is known as a {\em Legendre transformation}.
Successive application allows
one to calculate all thermodynamic potentials
from the grand potential $\Omega$
and hence, ultimately, from the partition function $Z$.
\label{born}
}
\end{figure}

Like the partition function, 
thermodynamic potentials
are auxiliary quantities used to facilitate
calculations. 
One example is the
(generalized) {\em grand potential}
\begin{equation}
\Omega(T,l^a,h^b):=-{1\over\beta}\ln Z
\quad,
\end{equation}
related to the internal energy $U$ via
\begin{equation}
\Omega=U-TS+ \sum_a l^a g_a
\quad.
\end{equation}
Its differential
\begin{equation}
{d}\Omega=-S{d} T+ \sum_a g_a{d} l^a- \sum_b m_b{d} h^b
\end{equation}
shows that 
$S$, $g_a$ and $m_b$ can be obtained from the grand potential 
by partial differentiation; e.g.,
\begin{equation}
S=-\left({\partial\Omega\over\partial T}\right)_{l^a,h^b}
\quad,
\end{equation}
where the subscript means that the partial derivative is to be taken
at fixed $l^a,h^b$.
In addition to the grand potential
there are many other thermodynamic potentials:
Their definition and properties are best summarized
in a Born diagram (Fig. \ref{born}).
In a given physical situation it is most convenient to work with that
potential which depends on the variables being controlled
or measured in the experiment.
For example, if a chemical reaction takes place at
constant temperature and pressure (controlled variables
$T$, $\{m_b\}=\{p\}$), 
and the observables of
interest are the particle numbers of the various
reactants (measured variables $\{g_a\}=\{N_i\}$)
then the reaction is most conveniently described by the free enthalpy
$G(T,N_i,p)$.
\def\epsfsize#1#2{0.7#1}

When a large system is physically divided into several subsystems then
in these subsystems 
the thermodynamic variables generally take values that 
differ from those of the total system.
In the special case of a {\em homogeneous system}
all variables of interest can be classified either as
extensive --varying proportionally to the volume of the respective
subsystem--
or intensive --remaining invariant under the subdivision of
the system.
Examples for the former are the volume itself,
the internal energy or the number of
particles; whereas amongst the latter are the 
pressure, the temperature or the chemical potential.
In general, if a thermodynamic variable is extensive then its
conjugate is intensive, and vice versa.
If we assume that the temperature and the $\{l^a\}$ are intensive, while
the $\{h^b\}$ and the grand potential are extensive,
then
\begin{equation}
\Omega_{\rm hom}(T,l^a,\tau h^b)=\tau\cdot \Omega_{\rm hom}(T,l^a,h^b)
\quad\forall\quad\tau>0
\end{equation}
and hence
\begin{equation}
\Omega_{\rm hom}=- \sum_b m_b h^b
\quad.
\end{equation}
This implies the {\em Gibbs-Duhem relation}
\begin{equation}
S{d} T- \sum_a g_a{d} l^a- \sum_b h^b{d} m_b=0
\quad.
\end{equation}
For an ideal gas in the grand canonical ensemble, for instance, we have
the temperature $T$ and the chemical potential
$\{l^a\}= \{-\mu\}$ intensive, whereas the volume $\{h^b\}= \{V\}$ and 
the grand potential $\Omega$ are extensive; hence
\begin{equation}
\Omega_{\rm i.gas}(T,\mu,V) = - p(T,\mu)\, V
\quad.
\end{equation}

\subsection{Correlations}

Arbitrary expectation values $\langle A\rangle_\rho$ in the 
macrostate (\ref{cl_macro}) or (\ref{qu_macro}), respectively,
depend on 
the Lagrange multipliers $\{\lambda^a\}$ as well as --possibly-- on other
parameters $\{h^b\}$.
If the Lagrange multipliers  vary infinitesimally while the $\{h^b\}$ are
held fixed, the
expectation value $\langle A\rangle_\rho$ changes according to
\begin{equation}
{d}\langle A\rangle_\rho
= -\sum_a \langle\delta G_a;A\rangle_\rho {d}\lambda^a
\quad.
\end{equation}
Here $\langle;\rangle_\rho$ is 
the {\em canonical correlation function}
with respect to the state $\rho$:
\begin{equation}
\langle A;B\rangle_\rho :=
\int d{\bf{\pi}}\,\rho({\bf{\pi}})A({\bf{\pi}})^* B({\bf{\pi}})
\end{equation}
in the classical case or
\begin{equation}
\langle A;B\rangle_\rho :=
\int_0^1{d}\nu\,\mbox{tr}
\left[\hat{\rho}^\nu \hat{A}^\dagger \hat{\rho}^{1-\nu} \hat{B}\right]
\end{equation}
in the quantum case, respectively.
The observable $\delta G_a$ is defined as
\begin{equation}
\delta G_a:=G_a-\langle G_a\rangle_\rho
\quad.
\end{equation}
The {\em correlation matrix}
\begin{equation}
C_{ab}:=
\langle{\delta G_a;\delta G_b}\rangle_\rho
= -\left({\partial g_b\over\partial\lambda^a}\right)_{\lambda,h}
=\left(
{\partial^2\over\partial\lambda^a\partial\lambda^b}\ln Z
\right)_{\lambda,h}
\end{equation}
thus relates infinitesimal variations of $\lambda$ and $g$:
\begin{equation}
{d} g_b = -\sum_a {d} \lambda^a C_{ab}
\quad,\quad
{d} \lambda^a = -\sum_b {d} g_b (C^{-1})^{ba}
\quad.
\end{equation}
The subscripts $\lambda, h$ of the partial derivatives indicate that 
they must be taken with all other $\{\lambda^a\}$ and all $\{h^b\}$
held fixed.
Returning to our example of the ideal quantum gas, we immediately
obtain from (\ref{occupation}) the correlation of occupation numbers
\begin{equation}
\langle{\delta N_i;\delta N_j}\rangle_\rho
= - {\partial n_j\over
\partial\alpha^i}
= \delta_{ij}\, n_i (1\pm n_i)
\quad.
\end{equation}

\section{Linear Response}

\subsection{Liouvillian and Evolution}

The dynamics 
of an expectation value $\langle A\rangle_\rho$ is
governed by the equation of motion
\begin{equation}
{d\langle A\rangle_\rho\over dt}=
\langle i{\cal L}A\rangle_\rho
+ \left\langle {\partial A\over\partial t}\right\rangle_\rho
\quad.
\end{equation}
Here we have allowed for an explicit time-dependence of
the observable $A$. Classically, the 
{\em Liouvillian} ${\cal L}$ takes the Poisson bracket
with the Hamilton function $H({\bf\pi})$,
\begin{equation}
i{\cal L}=\sum_j \left(
{\partial H\over\partial P_j}\,{\partial\over\partial Q^j}
-{\partial H\over\partial Q^j}\,{\partial\over\partial P_j}
\right)
\end{equation}
in canonical coordinates ${\bf \pi}=\{Q^j,P_j\}$;
whereas in the quantum case it takes the commutator with
the Hamilton operator $\hat{H}$,
\begin{equation}
i{\cal L}= (i/\hbar)\, [\hat{H},*]
\quad.
\end{equation}
An observable $A$ for which $i{\cal L}A+\partial A/\partial t=0$ 
is called a {\em constant of the motion;}
a state $\rho$ for which 
${\cal L}\rho=0$ is called {\em stationary}.
Only for a stationary $\rho$ the Liouvillian
is Hermitian with respect to 
the canonical correlation function,
\begin{equation}
\langle A;{\cal L}B\rangle_\rho = \langle {\cal L}A;B\rangle_\rho
\quad\forall\, A,B
\quad.
\end{equation}

The {\em evolver} ${\cal U}$ is defined as
the solution of the differential equation
\begin{equation}
{\partial\over\partial t}{\cal U}(t_0,t)={i}
{\cal U}(t_0,t){\cal L}
\end{equation}
with initial condition ${\cal U}(t_0,t_0)=1$.
As long as the Liouvillian ${\cal L}$ is not explicitly
time-dependent, the solution has the simple exponential form
\begin{equation}
{\cal U}(t_0,t)=\exp[{i}(t-t_0){\cal L}]
\quad;
\end{equation}
however, we shall not assume this
in the following.
The evolver determines --at least formally-- the evolution of expectation
values via
\begin{equation}
\langle A\rangle_\rho (t)=\langle {\cal U}(t_0,t) A\rangle_\rho (t_0)
\quad.
\end{equation}
Multiplication with a step function 
\begin{equation}
\theta(t-t_0) = \left\{
\begin{array}{ll}
0 &:\,t\le t_0 \\
1 &:\,t>t_0
\end{array}
\right.
\end{equation}
yields
the so-called {\em causal evolver}
\begin{equation}
{\cal U}_< (t_0,t):={\cal U} (t_0,t) \cdot\theta(t-t_0)
\end{equation}
(where `$<$' symbolizes `$t_0<t$') which
satisfies another differential equation
\begin{equation}
{\partial\over\partial t}{\cal U}_< (t_0,t)=
{i}{\cal U}_< (t_0,t){\cal L} + \delta(t-t_0)
\quad.
\end{equation}
If a (possibly time-dependent) perturbation is added to the
Liouvillian,
\begin{equation}
{\cal L}^{(V)}:={\cal L}+ {\cal V}
\quad,
\end{equation}
then the perturbed causal evolver 
${\cal U}^{(V)}_<$ is related to the unperturbed ${\cal U}_<$
by an integral equation
\begin{equation}
{\cal U}^{(V)}_< (t_0,t)=
{\cal U}_< (t_0,t)+\int_{-\infty}^\infty {d}t'\,
{\cal U}^{(V)}_< (t_0,t') \,{i}{\cal V}(t')\,
{\cal U}_< (t',t)
\quad.
\label{pert_series}
\end{equation}
Iteration of this integral equation
--re-expressing the ${\cal U}^{(V)}_< (t_0,t')$ in the integrand
in terms of another sum of the form (\ref{pert_series}),
and so on-- yields an infinite series, the terms being of
increasing order in ${\cal V}$.
Truncating this series after the term of order ${\cal V}^n$
gives an approximation to the exact causal evolver in
{\em $n$-th order perturbation theory}.

\subsection{Kubo formula}

The Kubo formula describes the response of a system to  
weak time-dependent external fields $\phi^\alpha(t)$.
Before $t=0$ the external fields are zero and
the system is assumed to be in an initial equilibrium state
\begin{equation}
\rho(0)=
{1\over Z}\exp\left(-\sum_a\lambda^a G_a[0]\right)
\end{equation}
characterized by some set
$\{G_a[0]\}$ of constants of the motion at zero field
(and with the a priori distribution $\sigma$ taken to be uniform).
Then the external fields are switched on:
\begin{equation}
\phi^\alpha(t)=\left\{
\begin{array}{ll}
0 &:\,t\le 0 \\
\phi^\alpha(t) &:\,t>0
\end{array}
\right.
\quad.
\end{equation}
How does an arbitrary expectation value $\langle A\rangle(t)$
evolve in response to this external perturbation?
The general solution is
\begin{equation}
\langle A\rangle (t)=
\langle{\cal U}^{[\phi]}_< (0,t) A\rangle_0
\quad,
\end{equation}
where $\langle\rangle_0$ stands for the expectation value in the
initial equilibrium state $\rho(0)$.
We assume that the observable $A$ does not depend explicitly
on time or on the fields $\phi^\alpha(t)$. 
The Hamiltonian $H[\phi]$ and with it the Liouvillian ${\cal L}[\phi]$, 
on the other hand, generally do depend on the external fields. 
Provided the fields are sufficiently weak, 
the Liouvillian may be expanded linearly:
\begin{equation}
{\cal L}[\phi(t)]\approx {\cal L}[0] + 
\sum_\alpha \left.{\partial {\cal L}[\phi]\over
\partial \phi^\alpha}\right|_{\phi=0} \phi^\alpha(t)
\quad.
\end{equation}
The zero-field Liouvillian ${\cal L}[0]$ is assumed to be not 
explicitly time-dependent;
the linear correction to it generally is, and may be regarded as
a time-dependent perturbation ${\cal V}(t)$.
Application of first order time-dependent perturbation theory
then yields the evolver ${\cal U}^{[\phi]}_<$ in
terms of ${\cal V}(t)$ and the zero-field
evolver ${\cal U}_<$.
Assuming for simplicity that $\langle A\rangle_0=0$
we thus find
\begin{equation}
\langle A\rangle (t)=
\sum_\alpha
\int_{-\infty}^\infty {d}t'\,
\left\langle
{i}\left.{\partial{\cal L}[\phi]\over\partial \phi^\alpha}\right|_{\phi=0}
{\cal U}_< (t',t)A
\right\rangle_0 \phi^\alpha(t')
\quad.
\end{equation}
With the help of the mathematical identity (prove it!)
\begin{equation}
\langle{i}{\cal L}[\phi]B\rangle_0=
\sum_a
\langle{i}{\cal L}[\phi]G_a[0];B\rangle_0
\lambda^a
\quad\forall\,\,B
\end{equation}
we can also write
\begin{equation}
\langle A\rangle (t)=
\sum_\alpha
\int_{-\infty}^\infty {d}t'\sum_a
\lambda^a 
\left\langle{i}
\left.{\partial{\cal L}[\phi]\over\partial \phi^\alpha}\right|_{\phi=0}
G_a[0];{\cal U}_< (t',t)A
\right\rangle_0 \phi^\alpha(t')
\quad.
\end{equation}
In general, the constants of the motion 
depend explicitly on the external fields.
They satisfy
\begin{equation}
{\cal L}[\phi]G_a[\phi]=0
\quad\forall\,\phi
\quad,
\end{equation}
yet generally ${\cal L}[\phi']G_a[\phi]\ne 0$
for $\phi'\ne\phi$.
Together with the Leibniz rule this implies
\begin{equation}
\left.{\partial{\cal L}[\phi]\over\partial \phi^\alpha}\right|_{\phi=0}
G_a[0]=
-{\cal L}[0]
\left.{\partial G_a[\phi]\over\partial \phi^\alpha}\right|_{\phi=0}
\quad,
\end{equation}
which we use to obtain
\begin{equation}
\langle A\rangle (t)=
-\sum_\alpha \int_{-\infty}^\infty {d}t'\sum_a
\lambda^a 
\left\langle
{i}{\cal L}[0]
\left.{\partial{G_a}[\phi]\over\partial \phi^\alpha}\right|_{\phi=0};
{\cal U}_< (t',t)A
\right\rangle_0 \phi^\alpha(t')
\quad.
\end{equation}
The right-hand side of this equation has the structure of a convolution,
so in the frequency representation we obtain an ordinary product
\begin{equation}
\langle A\rangle (\omega)
=
\sum_\alpha
\chi_\alpha^A(\omega)\phi^\alpha(\omega)
\quad.
\end{equation}
The coefficient
\begin{equation}
\chi_\alpha^A(\omega)
=
-\sum_a \lambda^a \int_0^\infty dt\,\exp(i\omega t)
\left\langle
{i}{\cal L}[0]
\left.{\partial{G_a}[\phi]\over\partial \phi^\alpha}\right|_{\phi=0};
A(t)
\right\rangle_0
\label{kubo}
\end{equation}
with $A(t):={\cal U}_< (0,t)A$ is called the
{\em dynamical susceptibility}.
The above expression for the dynamical susceptibility
is known as the {\em Kubo formula}.

\subsection{Example: Electrical conductivity}

The conductivity $\sigma^{ik}(\omega)$
determines the linear response of the current
density $\vec{j}$ to a (possibly time-dependent) 
homogeneous external electric field $\vec{E}$.
We identify
\begin{equation}
\phi^\alpha\to E_i\quad,\quad A\to  j^k\quad,\quad
\chi_\alpha^A(\omega)\to\sigma^{ik}(\omega)
\quad.
\end{equation}
Since a conductor is an open system with the number of electrons
fixed only on average, its initial state must be described by
a grand canonical ensemble: $\{G_a[\phi]\}\to\{H[\vec{E}],N\}$,
with associated Lagrange parameters $\{\lambda^a\}\to\{\beta,-\beta\mu\}$.
In principle, the formula for the conductivity then contains
both $\partial H/\partial E_i$ and $\partial N/\partial E_i$;
but the latter vanishes, and there remains only 
\begin{equation}
{\partial H\over\partial E_i}=-e Q^i
\quad,
\end{equation}
with $Q^i$ denoting the $i$-th component of 
the position observable and $e$ the electron charge.
We use the general formula (\ref{kubo}) for the
susceptibility to obtain
\begin{equation}
\sigma^{ik}(\omega)= e\beta \int_0^\infty dt\,\exp(i\omega t)
\langle{i}{\cal L}[0] Q^i; j^k(t)\rangle_0
\quad.
\end{equation}
The current density is related to the velocity
$V^k$ by
\begin{equation}
j^k=en V^k
\quad,
\end{equation}
where $n$ is the number density of electrons.
Furthermore, ${i}{\cal L}[0] Q^i=V^i$.
Hence the conductivity is proportional to the
velocity-velocity correlation:
\begin{equation}
\sigma^{ik}(\omega)= e^2 n\beta \int_0^\infty dt\,\exp(i\omega t)
\langle V^i; V^k(t)\rangle_0
\quad.
\end{equation}
This result is rather intuitive.
In a
dirty metal or semiconductor, for instance,
the electrons will often scatter off impurities,
thereby changing their velocities.
As a result, the velocity-velocity correlation function will decay rapidly,
leading to a small conductivity.
In a clean metal with fewer impurities, on the other hand, 
the velocity-velocity correlation function will decay more slowly,
giving rise to a correspondingly larger conductivity.
\end{document}